\documentclass[preprint, 5p]{elsarticle}

\usepackage{amsmath,amsfonts,mathrsfs}
\usepackage{multirow, diagbox, makecell}
\usepackage{subcaption, caption}
\usepackage{color, changes, soul}
\usepackage{hyperref}
\usepackage{lineno}
\usepackage{float}

\journal{Optics and Lasers in Engineering}

\biboptions{sort&compress}

\begin{document}
	
	\begin{frontmatter}
	
		\title{Robust Phase Retrieval with Green Noise Binary Masks}
		
		\author[mainaddress]{Qiuliang Ye}
		\author[mainaddress]{Yuk-Hee Chan}
		\author[secondaryaddress,thirdaddress]{Michael G. Somekh}
		\author[mainaddress]{Daniel P.K. Lun\corref{correspondingauthor}}
		\cortext[correspondingauthor]{Corresponding author}
		\ead{enpklun@polyu.edu.hk}

		\address[mainaddress]{Department of Electronic and Information Engineering, The Hong Kong Polytechnic University, Hong Kong SAR, People’s Republic of China}
		\address[secondaryaddress]{Center for Nanophotonics, Shenzhen University, Shenzhen PRC 518060, People’s Republic of China}
		\address[thirdaddress]{Faculty of Engineering, University of Nottingham, Nottingham NG7 2RD, United Kingdom}

		\begin{abstract}
			Phase retrieval with pre-defined optical masks can provide extra constraints and thus achieve improved performance. Recent progress in optimization theory demonstrates the superiority of random masks in enhancing the accuracy of phase retrieval algorithms. However, traditional approaches only focus on the randomness of the masks but ignore their non-bandlimited nature. When using these masks for phase retrieval, the intensity measurements contain many significant high-frequency components that the phase retrieval algorithm cannot take care of and thus leads to degraded performance. Based on the concept of digital halftoning, this paper proposes a green noise binary masking scheme that can significantly reduce the high-frequency contents of the masks while fulfilling the randomness requirement. The resulting intensity measurements will contain data concentrated in the mid-frequency band and around zero frequency areas which can be fully utilized in the phase retrieval optimization process. Our experimental results show that the proposed green noise binary masking scheme consistently outperforms the traditional ones when using in binary coded diffraction pattern phase retrieval systems.
		\end{abstract}
	
	
	\end{frontmatter}


	\section{Introduction} \label{sec:1}
		
		Many measurement systems can only detect the intensity of the optical wave field.   Phase retrieval, as the name suggests, aims to reconstruct a complex-valued signal from  intensity-only measurements \cite{Gerchberg1972APA, Fienup:82, Miao1999ExtendingTM, rodenburg2008ptychography, XU201996, SHEN201854, GUO201816}. The phase retrieval problem is encountered in several disciplines, including crystallography, optical imaging, astronomy, X-ray, and electronic imaging.  There has been considerable recent progress in phase retrieval algorithms due to the advent of modern optimization theories \cite{Shechtman2015PhaseRW}. In particular, the so-called coded diffraction pattern (CDP) phase retrieval algorithm adopts pre-defined optical masks to improve the reconstruction performance with extra constraints \cite{candes2015code, candes2015phase}. Specifically, the algorithm reconstructs a signal  $\mathbf{U}\in \mathbb{C}^n$ (complex-valued) based on the Fourier intensity measurements $y_i = |\mathcal{F}(\mathbf{I}_i\circ \mathbf{U})|^2, i = 1,\dots, M$, where $M$, $\mathbf{I}$ and $\circ$ denote the number of measurements, pre-defined optical masks and elementwise multiplication, respectively; and $ \mathcal{F}  $ is the Fourier transform. In practice, the pre-defined masks can be implemented using a spatial light modulator (SLM) or digital micromirror device (DMD) \cite{Falldorf2010PhaseRB, Horisaki2014SingleshotPI, Horisaki2015ExperimentalDO,  Zheng2017DigitalMD}. The number of measurements is related to the complexity of the signal and the masks. For instance, the single-shot lensless phase retrieval methods in \cite{Horisaki2014SingleshotPI, Horisaki2015ExperimentalDO} require only a single measurement, but it can reconstruct only sparse objects.  On the other hand, using multiple-level phase masks can also reduce the number of measurements. However, they require optical devices that can perform accurate multi-level phase modulation for their implementation. The cost of these devices is one concern; the error of these devices due to the global grayscale-phase mismatch and spatial non-uniformity is another. If binary masks are used, it is empirically shown in \cite{candes2015code, candes2015phase} that at most $ 6 $ measurements are required for exact recovery via convex programming. Thus, the number of measurements is not excessive. Binary masks can provide accurate reconstruction as the two levels can be reliably reproduced; these can be realized with  both  DMDs and  SLMs.  For this reason, binary masks are used in this work. 
		
		The importance of the randomness of the masks is emphasized in some research papers to ensure good phase retrieval performance \cite{candes2015code, candes2015phase}. Traditionally, white noise masks are commonly used  since  randomness is guaranteed. However, the Fourier spectrum of white noise masks is non-bandlimited in theory. When using them for phase retrieval, the intensity measurements can have many significant high-frequency components located beyond the $ 0 $th diffraction order (an example is shown in Figure \ref{fig:Fourierintensity}(a), the yellow circle denotes the boundary of the 0th diffraction order). In general, we always focus on the $ 0 $th diffraction order such that the data outside the $ 0 $th diffraction order are not detected. Even if we also detect the data outside the $ 0 $th diffraction order, they will still be ignored in the phase retrieval optimization process. This is because  traditional CDP phase retrieval algorithms assume the masked data as a grid of sampled data so that its spectrum will be periodic based on the sampling theory. The algorithm can only handle the frequency components within the Nyquist frequency (i.e., the red square in Figure \ref{fig:Fourierintensity}(a)). Therefore, the high-frequency components located beyond the Nyquist frequency of the system will be ignored by the phase retrieval algorithm. It is equivalent to filtering out the high-frequency components of the intensity measurements, and this introduces errors to the optimization process.  In this paper, we base on our previous work \cite{fung2010green} on digital halftoning to propose a green noise optical masking scheme for phase retrieval. The proposed scheme allows the optical masks to have the green noise property, which means that the energy of the masks is concentrated in a ring shape mid-frequency region. It allows the intensity measurements to also have the data concentrated in the same region (plus those around the zero frequency, an example is shown in Figure \ref{fig:Fourierintensity}(b)).  Thus, the amount of the filtered high-frequency components can be reduced. Halftoning is a process  commonly used in digital printers to simulate shades of gray by varying the size of tiny black dots. Recently, we apply the multiscale error diffusion (MED) technique to the halftoning process so that we can control the black dot patterns to have the green noise property \cite{fung2010green}. Such a technique can also be used in optical mask generation. One advantage of the scheme is that we can generate the binary masks of different densities (the ratio of the number of ones to zeros in the masks)  while maintaining the green noise property. The density of the masks needs to be carefully chosen to avoid the saturation problem when taking the measurements. By using the proposed green noise masking scheme, the error of the reconstructed phase images can be reduced compared with using the traditional white noise masks. It has been verified with our experimental results  in Section \ref{sec:4}.

	\section{Phase Retrieval with Coded Diffraction Patterns} \label{sec:2}
			
		The objective of the CDP phase retrieval algorithm \cite{candes2015code, candes2015phase} can be described by Eq. \eqref{CDP}.
			\begin{equation}
				\label{CDP}
				\begin{aligned}
					\text { Find } \mathbf{U} \in \mathbb{C}^{n} \quad \text { s.t. } \quad \mathbf{y}_{i}=\left|\mathcal{F}\left(\mathbf{I}_{i} \circ \mathbf{U}\right)\right|^{2}, i=1, \ldots, M,
				\end{aligned}
			\end{equation}
		where $ \mathbf{y}_i $ is a Fourier intensity measurement and $\circ $ denotes the elementwise multiplication. The use of the optical masks $ \mathbf{I} $ provides the constraints to reduce the ill-posedness of the problem. It has been shown in \cite{candes2015code, candes2015phase} that $ \mathbf{U} $ can be perfectly reconstructed from noiseless measurements $ \mathbf{y} $ with high probability given $ \mathbf{I} $ are random and $ M $ is sufficiently large. However, in a coherent imaging system, speckle noise inherent in the laser and shot noise due to the limited photon measurements are prevalent. They bring errors to the Fourier intensity measurements and corrupt the reconstructed phase. The maximum a-posterior (MAP) method is often used to ensure the performance of the estimation. It is applied for estimating the maximum point of an unknown quantity with the help of a posterior distribution. To further improve the optimization performance, different regularization terms are added to the optimization cost function. In this work, we adopt the total variation (TV) regularization term to smooth out the noise while preserving the image edges \cite{chang2018total}. Combining with the MAP formulation, we adopt the following TV-MAP approach \cite{chang2018total} for the implementation of the CDP phase retrieval algorithm in this paper: 
		
		\begin{equation}
			\label{TV-MAP}
			\begin{array}{c}
				\min \limits_{\mathbf{U} \in \mathbb{C}^{n}} \alpha\|\nabla \dot{\mathbf{U}}\|_{1}+\frac{1}{2} \sum_{i}\left(\left|\dot{\mathbf{g}}_{i}\right|^{2}-\dot{y}_{i} \log \left|\dot{\mathbf{g}}_{i}\right|^{2}\right) \\
				\text { s.t. } \dot{\mathbf{g}}_{i}=\left|D F T\left(\mathbf{I}_{i} \circ \dot{\mathbf{U}}\right)\right|^{2}, i=1, \cdots, M
			\end{array}
		\end{equation}
			
		In Eq.\eqref{TV-MAP}, $ \nabla $ is the gradient operator, $ \|\cdot\|_1 $ refers to the $ \ell_1 $ norm and DFT refers to the discrete Fourier transform; and $ \dot{\mathbf{y}} $ represents the discrete Fourier intensity measurements  which are the sampled version of $ \mathbf{y} $. Similarly, $ \dot{\mathbf{I}} $ and $ \dot{\mathbf{U}} $ are the samples of $ \mathbf{I} $ and $ \mathbf{U} $, respectively. The first term of Eq. \eqref{TV-MAP} is the TV regularization term that has been popularly used in noise removal. It is based on the principle that images with spurious details have higher total variation. Reducing the total variation of the measurement can remove the unwanted details. The second term of Eq. \eqref{TV-MAP} is the MAP data fidelity term that evaluates the  performance of the estimation. Minimizing these two terms can obtain an estimate that is close to the original $ \dot{\mathbf{U}} $ while reducing the spurious detail due to noise. Furthermore, the optimal solution allows discontinuities along the curves; therefore, edges can be preserved in the restored image. The regularization strength is controlled by the constant $ \alpha $. It is usually selected empirically according to the complexity of the signal. Similar to \cite{candes2015code, candes2015phase}, Eq. \eqref{TV-MAP} can be solved via the Alternating Direction Method of Multipliers (ADMM) algorithm \cite{Boyd2011DistributedOA}. 
		
\begin{figure}[H]		
			\centering\includegraphics[width=\linewidth]{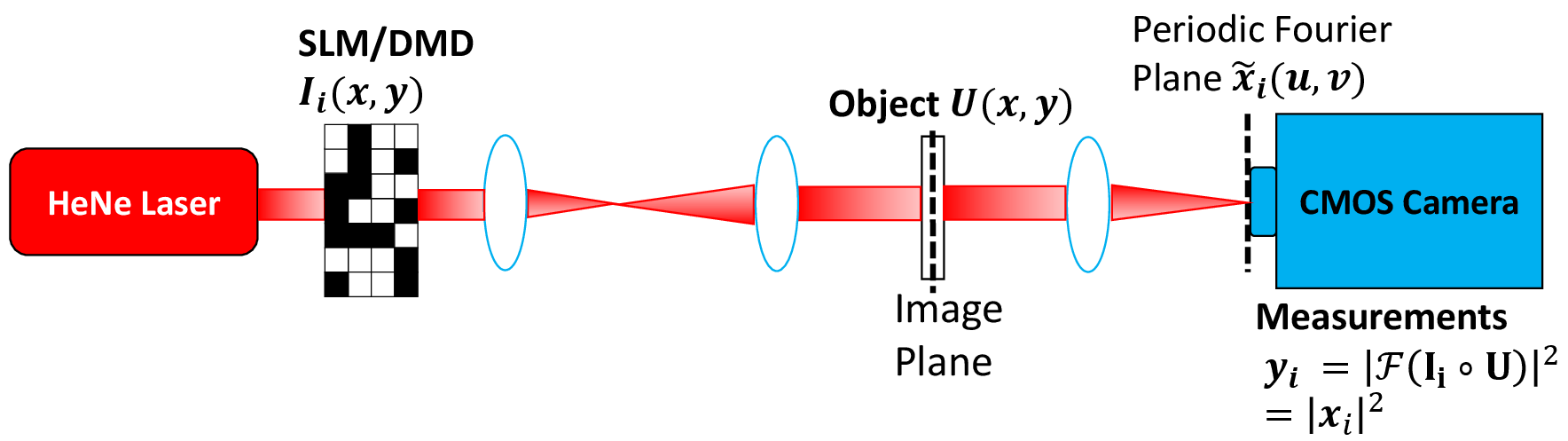}
			\caption{Experimental setup of the  CDP phase retrieval system.  }
			\label{fig:setup}
		\end{figure}

		As mentioned in Section \ref{sec:1}, binary masks are used in this work.  Both SLM and DMD can be used to implement  $\mathbf{I}$. When SLM is used, $\mathbf{I} $ can be implemented as a set of binary phase masks such that $ \mathbf{I} \in\{1, -1\} $. If DMD is used, $\mathbf{I} $ becomes a set of binary amplitude masks such that $\mathbf{I} \in \{0, 1\} $. Figure \ref{fig:setup} shows a typical experimental setup for implementing the  CDP phase retrieval system. It is also adopted in our experiments. In this optical system, the light from the HeNe laser is projected onto a  SLM or DMD board which implements the random masks. The masked light is projected onto an object  placed on the image plane. The light then goes through a lens so that the image captured by the CMOS camera (placed on the back focal plane of the lens) is the Fourier transform intensity $ \mathbf{y}_{i} $ of the transmittance function of the object $ \mathbf{U} $ multiplied with the $ i $-th mask $ \mathbf{I}_i $. 
		
\begin{figure}[H]		
			\centering
			\includegraphics[width=\linewidth]{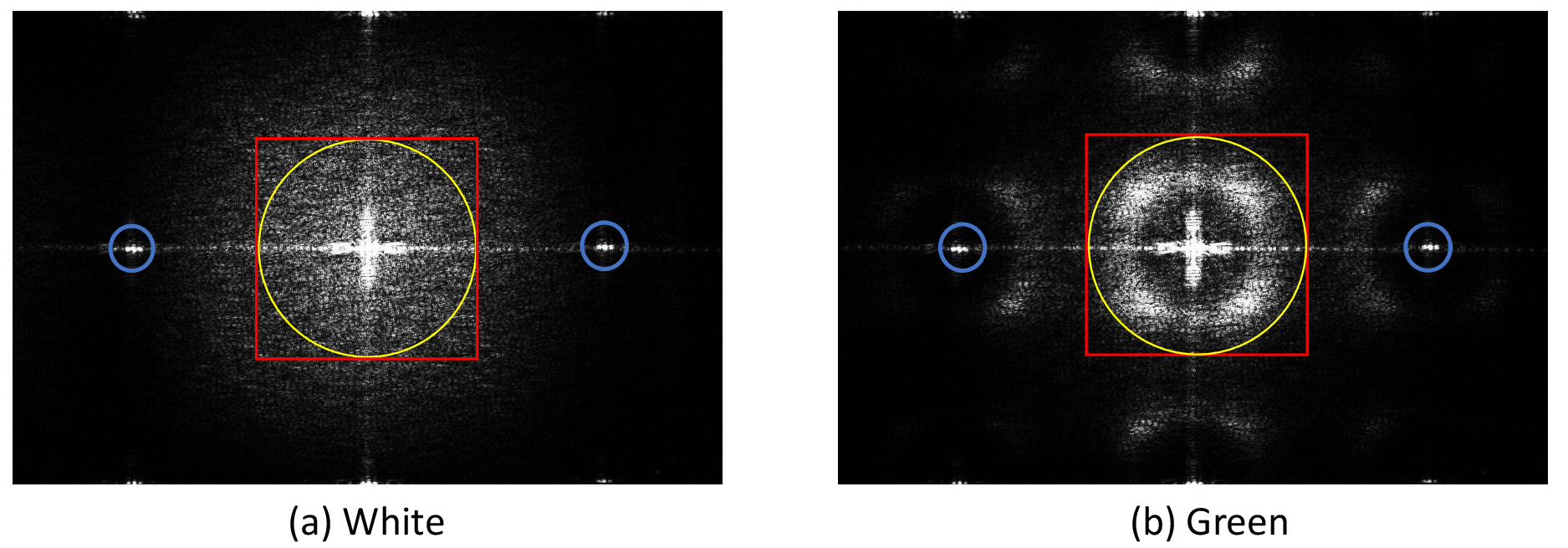}
			\caption{ Fourier intensity measurements of a USAF chart multiplied with (a) a white noise mask and (b) a green noise mask. The contrast of the images is adjusted to visualize the small coefficients. The yellow circle denotes the $ 0 $th diffraction order. The blue circles represent the central areas of the $ 1 $st diffraction order. The red square denotes the region where the phase retrieval algorithm will consider. The white noise mask introduces many high-frequency data that are beyond the red square. They will be ignored by the phase retrieval algorithm. The green noise mask allows the intensity measurement to have data concentrated in the mid-frequency region. They will be fully utilized by the phase retrieval algorithm.}
			\label{fig:Fourierintensity}
		\end{figure}

		As is indicated in \cite{Ding2014MicroscopicLW, Pan_2018},  $ \mathbf{y}_{i} $ appear as a kind of diffraction pattern  as shown in Figure \ref{fig:Fourierintensity}.  It should be noted that $ \mathbf{y}_{i} $ is obtained from the Fourier transform of the multiplication of $ \mathbf{I}_i $ and $ \mathbf{U} $ in spatial domain. This is equivalent to the convolution of $ \mathcal{F}(\mathbf{I}_i) $ and $ \mathcal{F}(\mathbf{U}) $ in frequency domain. However,  $ \mathcal{F}(\mathbf{I}_i) $ is non-bandlimited if $\mathbf{I}_i $ is white. Thus, the convolution of $ \mathcal{F}(\mathbf{I}_i) $ and $ \mathcal{F}(\mathbf{U}) $ is also non-bandlimited. It is the reason why we can find many high-frequency data in Figure \ref{fig:Fourierintensity}(a). In this paper, we propose to use the green noise masks that have a ring-shaped frequency spectrum concentrated in the mid-frequency region shown in Figure \ref{fig:differentratio}. If the green noise masks are used,  the convolution of $ \mathcal{F}\left(\mathbf{I}_i\right)$ and $\mathcal{F}\left(\mathbf{U}\right) $ will also have the data concentrated in the  mid-frequency region  presented in Figure \ref{fig:Fourierintensity}(b) (since $ \mathcal{F}\left(\mathbf{U}\right) $ has large magnitude around zero frequency). It can be seen that the high-frequency components of the $ 0 $th diffraction order are much weaker than those using the white noise masks. 
				
			The above property of the green noise masks is important for solving Eq. \eqref{TV-MAP}. In general, we always focus on the $ 0 $th diffraction order such that the data outside the $ 0 $th diffraction order are not detected. Even if we also detect the data outside the $ 0 $th diffraction order,   they will still be ignored in the phase retrieval optimization process. As mentioned above, the reconstruction algorithm  is a discrete process. In each iteration, the samples of $ \mathbf{U} $, i.e. $  \dot{\mathbf{U}} $, are used to compute ${{\dot{\mathbf{g}}}_i=\left|DFT\left({\dot{\mathbf{I}}}_i\circ\dot{\mathbf{U}}\right)\right|}^2 $. Then the MAP loss function evaluates the difference between $ {\dot{\mathbf{g}}}_i $ and  the discrete intensity measurement $ \dot{\mathbf{y}} $. The difference is then used to update the estimation of $ \dot{\mathbf{U}} $.  The sampling period of $ {\dot{\mathbf{I}}}_i $ and $  \dot{\mathbf{U}} $ is $ d $, i.e. the distance of the SLM pixels or the micro-display units of the DMD. Therefore, the Nyquist frequency of this discrete system is  $ \frac{\lambda f}{2d} $ denoted by the red square in Fig \ref{fig:Fourierintensity} (coincides with the boundaries of the $ 0 $th diffraction order circled in yellow with radius $ \frac{\lambda f}{2d} $).  Based on the sampling theory, $ {\dot{\mathbf{g}}}_i $ is periodic, and only one period of the replicated spectrum, i.e. from $ -\frac{\lambda f}{2d} $ to $ \frac{\lambda f}{2d} $ will be used by the algorithm to compare with the same region of $ {\dot{\mathbf{y}}} $.  Therefore,  the high-frequency data of $ {\dot{\mathbf{y}}} $ beyond the red square will not be considered in the phase retrieval process. It is equivalent to filtering out these high-frequency data. 
		
		As mentioned above, the intensity measurements obtained using the traditional white noise masks will contain many significant high-frequency components. Some of them can even distribute outside the red square shown in Figure \ref{fig:Fourierintensity}(a). These data will not be considered in the phase retrieval process and thus degrades the performance.  In Figure \ref{fig:Fourierintensity}(b), the data of the intensity measurements with the green noise masks concentrates in the mid-frequency region (plus around the zero frequency) and are well within the red square. They will be fully utilized in the phase retrieval process.  We will verify the advantages of  the green noise masks by the experimental results in Section  \ref{sec:4}.

	\section{Green Noise Binary Masks} \label{sec:3}
	
		\subsection{Generation of the Green Noise Masks}
	
\begin{figure}[H]		
				\centering\includegraphics[width=\linewidth]{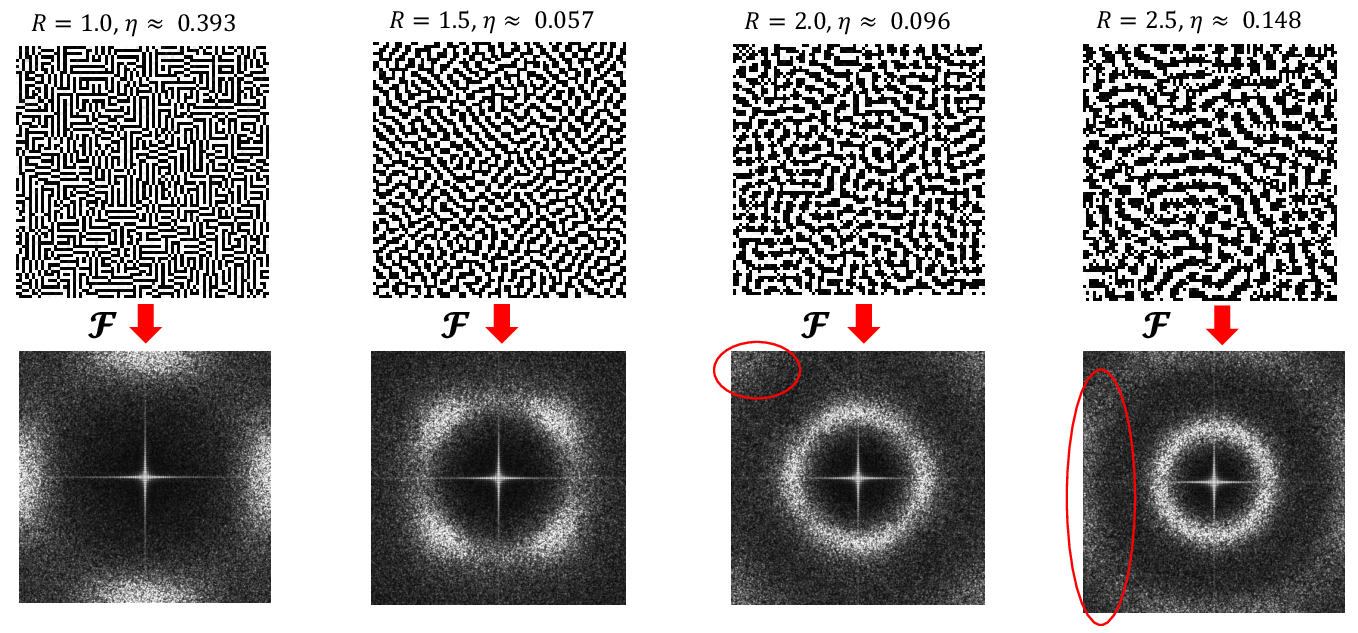}
				\caption{Spatial domain (upper), Fourier domain in log scale (lower) and energy concentration ratio of green noise masks of different $ R_1 $ with $ \sigma =0.5$. The parameter $ \eta $ is defined as the amount of energy with frequency greater than $ 0.8 $ of the maximum range. The regions circled in red indicate that the high-frequency data will become significant again when $ R_1 $ increases to a certain value.}
				\label{fig:differentratio}
			\end{figure}

\begin{figure}[H]		
				\centering\includegraphics[width=0.7\linewidth]{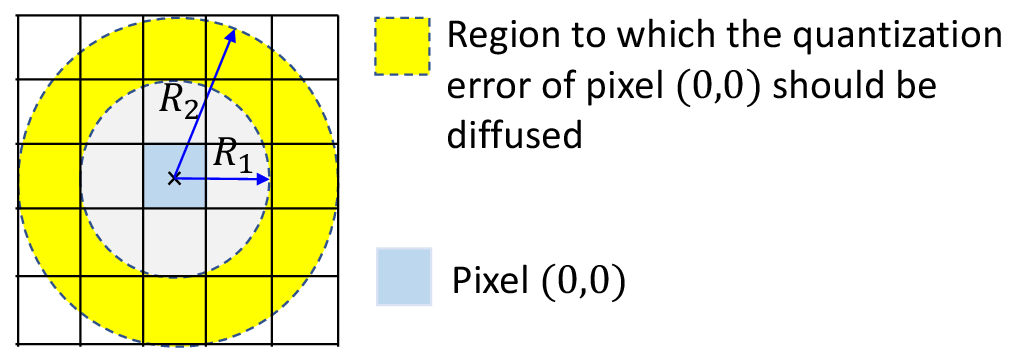}
				\caption{An illustration of the ring-shaped diffusion filter.}
				\label{fig:ringfilter}
			\end{figure}

			In Section \ref{sec:2}, we have explained why the non-bandlimited nature of the traditional white noise masks will lead to  significant high-frequency data that will be ignored in the phase retrieval optimization process and leads to degraded performance. In this paper, we propose to use the green noise masks with  energy concentrated in the mid-frequency region.  The DFT magnitude (in log scale) of some examples of green noise masks are shown in Figure \ref{fig:differentratio}. As can be seen in the figure, the energy of the masks concentrates in the ring in the middle of the Fourier spectrum. Thus, the high-frequency energy of the mask is significantly reduced. The green noise masks in Figure \ref{fig:differentratio} are designed based on a digital halftoning method called FMEDg that we proposed in \cite{fung2010green}. Digital halftoning is a process for turning a grayscale image into a binary image such that it can be physically printed \cite{Ulichney1987DigitalH}. When the input is a constant grayscale patch, the output of digital halftoning is a binary noise pattern composed of black $ (0) $ and white $ (1) $ dots. We make use of the digital halftoning method FMEDg in this research because it can generate binary noise patterns that fully fulfill the needs of a  CDP phase retrieval process. First, the output of digital halftoning is a set of black and white dots, which is the same as a binary mask. Second, the method can ensure the spatial distribution of the dots to be aperiodic, homogeneous and isotropic. Thus, it can fulfill the randomness requirement. More importantly, we can control the frequency response of the output noise pattern (from green to blue) and its density (i.e. the ratio of the number of ones to zeros in the masks). The algorithm starts with a mask with a constant value, for instance, $ 0.5 $. Then an iterative feature preserved multiscale error diffusion (FMED) approach is used to process each pixel in the mask \cite{Chan2004FeaturepreservingME}. More specifically, we select a not-yet processed pixel and quantize its value to be either $ 0 $ or $ 1 $. Then the quantization error is diffused to the pixel’s neighbors with a ring-shaped non-causal diffusion filter. This process repeats until all pixels are processed. One can refer to \cite{fung2010green} for the operation details. Since the method, which is based on FMED, can generate green noise patterns, the method is named as FMEDg. In that method, the diffusion filter plays a critical role in controlling the frequency response of the output pattern. Let us illustrate it through an example. Without loss of generality, we assume that pixel $ (0,0) $, which has the initial value of $ 0.5 $, is the pixel picked at a particular iteration. Assume it is quantized to $ 0 $, FMEDg will diffuse the quantization error, i.e. $ 0.5 $, to a ring-shaped region surrounding the pixel as shown in Figure \ref{fig:ringfilter}. The process is controlled by the following diffusion filter:
			
			\begin{equation}
				D(m, n)=\frac{A\left(m, n, R_{2}\right)-A\left(m, n, R_{1}\right)}{\left(R_{2}^{2}-R_{1}^{2}\right) \pi},
			\end{equation}
			where $ {A(m,n,R}_k) $ is a function such that,
			\begin{equation}
				A\left(m, n, R_{k}\right)=\left\{\begin{array}{ll}
					1, & \sqrt{m^{2}+n^{2}} \leq R_{k}, \ k=1,2 \text, R_{2}>R_{1} \\
					0, & \text { otherwise }
				\end{array}\right.
			\end{equation}
				
			More specifically, the error $ 0.5 $ will be multiplied by $D \left(m,n\right) $ and added to the pixel $ \left(m,n\right) $ if $ \left(m,n\right) $ is located inside the ring-shaped region determined by the outer radius $ R_2 $ and inner radius $ R_1 $. In \cite{fung2010green}, we show that by setting $ R_2=\sqrt{2}R_1 $, we can control the cluster size of the noise pattern to be approximately equal to $ R_1^2\pi\sigma $, where $ \sigma $ is the initial constant gray level of the mask and is also the final density of the noise pattern. The larger is the cluster size of the noise pattern, the lower the frequency response will be. And since the filter is isotropic, the frequency response of the noise pattern will appear as a ring-shaped pattern  shown in Figure \ref{fig:differentratio}. Thus, by following the FMEDg method, we can generate a green noise mask of any principal frequency and density $ \sigma $ by simply adjusting the parameter $ R_1 $ of the diffusion filter. 
			
		\subsection{Analysis of the Green Noise Mask Generated by the FMEDg Method} \label{sec:analysis}
		
			Figure \ref{fig:differentratio} shows some examples of the green noise masks of different $ R_1 $ generated by the FMEDg method (with the density $ \sigma $ fixed at $ 0.5 $). Here we would like to introduce another parameter $ \eta $ which indicates the energy of the mask in the high-frequency region of the spectrum. We define $ \eta $ as,
			\begin{equation}
				\eta(\mathbf{i})=\frac{\sum_{m, n} \mathbf{i}(m, n>0.8 N)}{\sum_{m, n} \mathbf{i}(m, n)},
			\end{equation}
			where $  \mathbf{i}=\left|DFT(\mathbf{I})\right|^2 $ and $ N $ is the total number of DFT coefficients. It can be seen in Figure \ref{fig:differentratio} that, as $ R_1 $ increases, the noise pattern in the mask becomes clustered and the ring shape frequency spectrum shrinks towards the center. Thus, the principal frequency of the noise pattern decreases. However, the high-frequency energy of the masks (indicated by $ \eta $) does not decrease as the principal frequency decreases. Rather, it first decreases but then increases as $ R_1 $ further increases. This is because as the ring shape frequency spectrum shrinks towards the center, another ring starts to appear at the high-frequency spectrum (circled in Figure \ref{fig:differentratio}). 
			
\begin{figure}[H]		
				\centering\includegraphics[width=0.8\linewidth]{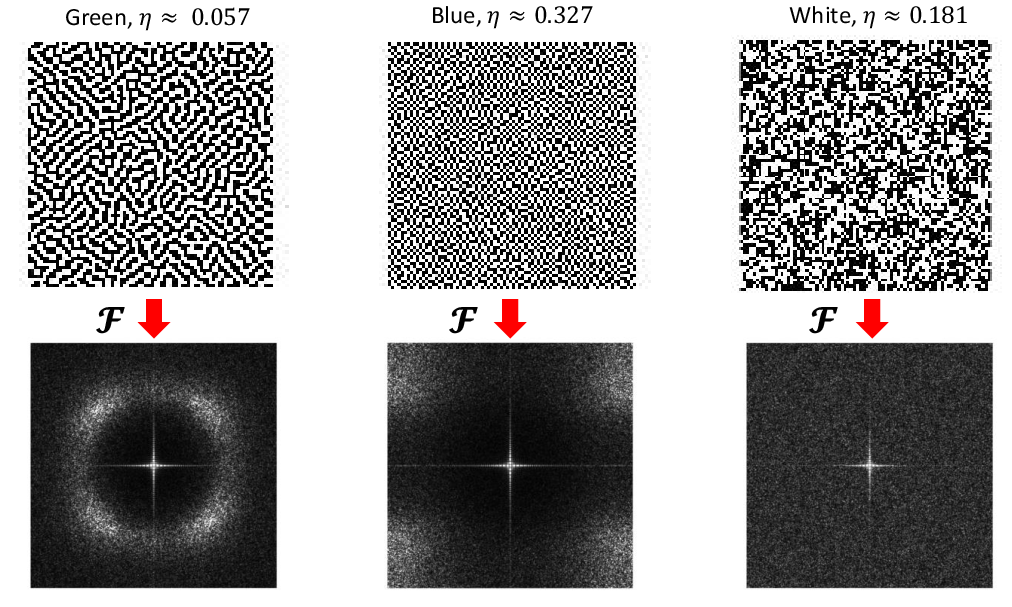}
				\caption{Spatial domain (upper), Fourier domain in log scale (lower) and energy concentration ratio of different kinds of binary masks.}
				\label{fig:differentmask}
			\end{figure}

\begin{figure*}[t]		
				\centering\includegraphics[width=0.7\linewidth]{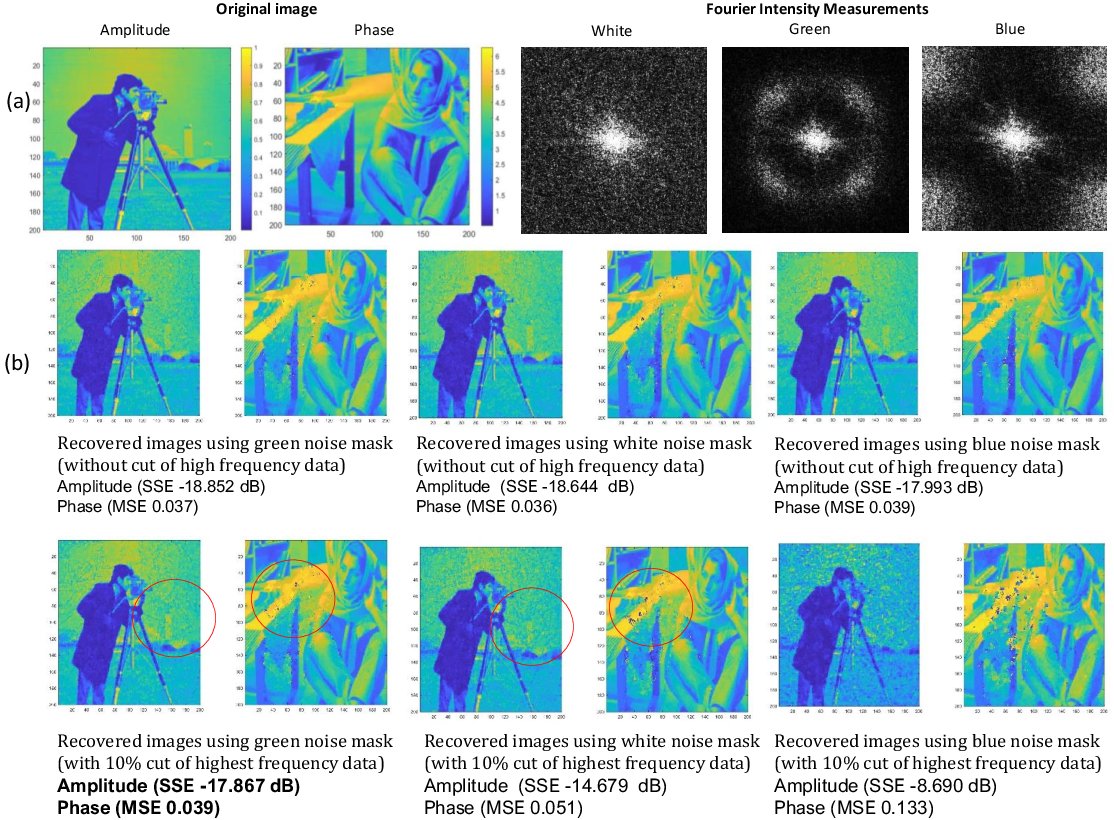}
				\caption{(a) Simulation images: Cameraman + Barbara, and the Fourier intensity measurements with white, green and blue noise masks (in log scale for visualizing the small coefficients). (b) Reconstruction results of using the green, blue and white noise masks with/without  $ 10\% $ of the highest frequency data setting as  $ 0 $. The same Poisson and Gaussian noises (noise variance $ = 15 $)  are added to the Fourier intensity measurements in each case. In this simulation, $ 4 $ Fourier intensity images ($M =  4 $) are used for each kind of noise mask. The performance difference of the green and white noise masks can be best observed in the regions circled in red. }
				\label{fig:simulation}
			\end{figure*}

			As it is explained above, the intensity measurement data of frequency higher than the Nyquist frequency of the system will be ignored in the phase retrieval optimization process. Although they are of very high frequencies, ignoring them can still significantly affect the performance. A  simulation is conducted to demonstrate their importance. We use two standard images namely Cameraman and Barbara (see Figure \ref{fig:simulation}(a)) as the amplitude and phase part of the input image. Then we multiply the image with the proposed green noise mask (with $ R_1=1.5 $ and $ \sigma=0.5 $), blue noise mask  and traditional white noise mask \cite{candes2015code}. The binary amplitude masks $ \mathbf{I} \in \{0, 1\} $ are used in the simulation. Figure \ref{fig:differentmask} shows the masks and their Fourier intensities. Some examples of the Fourier intensity measurements with these masks are shown in Figure {\ref{fig:simulation}}(a). For each intensity measurement, we set the $10\%$  highest frequency data to zero to demonstrate the effect when high-frequency data are ignored. Specifically, we create a binary mask with inner $90\%$ data set as $1$ while the outer $10\%$ data set as $0$. Then we multiply every intensity measurement with the binary mask to remove the high-frequency information. For each kind of mask, $ 4 $ Fourier intensity measurements ($ M = 4 $) are acquired and fed to the same CDP phase retrieval process described in Eq. \eqref{TV-MAP}. The number of measurements is not the main focus here. As long as $ M $ is not too small, we find in the simulation that the same conclusion can be drawn irrespective of the number of measurements. This is because having more masks cannot change the fact that $10\% $ highest frequency data is missing. If $ M $ is very small, the optimization process sometimes cannot converge. The performance is evaluated by two criteria: sum square error (SSE) of the amplitude and  mean square error (MSE) of the phase. They are defined as follows:
			
			\begin{equation}
				\begin{aligned}
					S S E_{\text {amplitude}} &=10 \log _{10} \frac{\sum_{i=0}^{N-1} \sum_{j=0}^{M-1}\left(\left|u_{i, j}\right|-\left|\tilde{u}_{i, j}\right|\right)^{2}}{\sum_{i=0}^{N-1} \sum_{j=0}^{M-1}\left(\left|u_{i, j}\right|\right)^{2}},\\
					M S E_{\text {phase}} &= \frac{1}{MN} \sum_{i=0}^{N-1} \sum_{j=0}^{M-1}\left(\angle u_{i, j}-\angle \tilde{u}_{i, j} - \phi \right)^{2},
				\end{aligned}
			\end{equation}
			where $ u $ and $ \widetilde{u} $ denote the original and reconstructed image, respectively. Both $ \angle u $ and $ \angle \tilde{u} $ are unwrapped to ensure the true phase difference is evaluated. A global phase shift $ \phi $ (a fixed constant) is further subtracted from the result to avoid the $ MSE_{phase} $ value to be amplified due to the bias in the estimated $ \angle \tilde{u} $. The values of $ SSE_{\text {amplitude}} $ are usually negative and expressed in dB. The smaller  the value of $ SSE_{\text {amplitude}} $, the better is the performance. The criterion is the same for $ MSE_{phase} $. To optimize the performance, we fine-tune the value of the regularization parameter $ \alpha $ in Eq. \eqref{TV-MAP} to ensure the best SSE is obtained in each case.
			
			The reconstruction performance using different masks is shown in Figure \ref{fig:simulation}(b). It can be seen that when all data are available, all masks give similar performance. When $ 10\% $ of highest frequency data are removed, the blue noise mask ($\eta=0.327$) gives the worst result since it has the highest high-frequency concentration. It will introduce many high-frequency data to the intensity measurements. The white noise mask ($ \eta=0.181 $)  commonly used in the literature also does not give a good result because its high-frequency concentration is still significant. On the other hand, the green noise mask ($ \eta=0.057 $)  gives the best performance both quantitatively and quantitatively. The degradation due to the removal of the high-frequency data is the least among  three masks. It can be best observed in the region circled in Figure \ref{fig:simulation}(b).  Table \ref{tablediffmask} further shows that the performance differences of the masking schemes are consistent at different noise levels. When $ 10\% $ of the highest frequency data are removed, the green noise mask always performs the best at all noise levels. This simulation has clearly indicated the importance of high-frequency data in the intensity measurements. In Section \ref{sec:4}, we further demonstrate the improvement of the proposed green noise masking scheme by applying it to two practical CDP phase retrieval systems. 
			
			\begin{table*}[t]
				\centering
				\small\addtolength{\tabcolsep}{-2pt}
				\captionsetup{width=.9\textwidth}
				\caption{A comparison of the performance when using different noise masks at different noise levels. In this simulation, $ 10\% $ of the highest frequency data are set as $ 0 $ to demonstrate the importance of high-frequency data. $ 4 $ Fourier intensity images ($ M=4 $) are used for each kind of noise mask.}
				
			    \begin{tabular}{|c|c|c|c|c|c|c| }
					\hline 
					\multirow{2}{*}{ \diagbox[width=4cm, height=1.15cm]{Gaussian \\ noise variance}{ Mask}}   & \multicolumn{2}{c|} { Green } & \multicolumn{2}{c|} { White } & \multicolumn{2}{c|} { Blue } \\
					\cline { 2 - 7 }  & \makecell[c]{ $ SSE_{amplitude} $\\$(\mathrm{dB})$} & \makecell[c]{ $ MSE_{phase} $}  & \makecell[c]{$ SSE_{amplitude} $\\$(\mathrm{dB})$} & \makecell[c]{$ MSE_{phase} $}  & \makecell[c]{$ SSE_{amplitude} $\\$(\mathrm{dB})$} & \makecell[c]{$ MSE_{phase} $} \\
					\hline $ 5 $ & $ \mathbf{-20.486}  $& $ \mathbf{0.015} $ & $ -16.243 $ & $ 0.017 $ & $ -10.792 $ & $  0.065  $\\
					\hline $ 10 $ & $ \mathbf{-19.274} $ & $ \mathbf{0.023} $ & $ -15.513 $ & $ 0.030 $ & $ -9.289 $ & $ 0.097 $ \\
					\hline $ 15 $ & $  \mathbf{-18.023} $ & $ \mathbf{0.040} $ & $ -14.449 $ & $ 0.049 $ & $ -8.443 $ & $ 0.130 $ \\
					\hline $ 20$ & $  \mathbf{-14.353}  $& $ \mathbf{0.073} $ & $  -13.585 $ & $  0.086 $ & $ -7.974 $ & $  0.172  $\\
					\hline $ 25 $ & $ \mathbf{-13.893} $ & $ \mathbf{0.104} $ & $ -12.464 $ & $ 0.122 $ & $ -7.661 $ & $ 0.224 $ \\
					\hline $ 30	 $ & $ \mathbf{-12.820} $ & $ \mathbf{0.149} $ & $ -11.422 $ & $ 0.173 $ & $ -7.187 $ & $ 0.273 $ \\
					\hline
				\end{tabular}
				\label{tablediffmask}
			\end{table*}
		
		\subsection{The Choice o f $ R_1 $ and $ \sigma $} \label{sec:choice}
		
			As mentioned above, the parameters $ R_1 $ and $ \sigma $ of FMEDg should be carefully chosen. We use the following simulation to show how they will affect the performance of the phase retrieval process. First, we use the FMEDg method to generate a few green noise masks of different $ R_1 $ and $ \sigma $. The binary amplitude masks $ \mathbf{I} \in \{0, 1\} $ are used in the simulation. The sizes of the test images and masks are both $ 200 \times 200 $ pixels. For each $ R_1 $ and $ \sigma $, $ 4 $ green noise masks are generated and multiplied with the images. The DFTs of the resulting images are computed and  $ 4  $  intensity measurements are obtained. Note that the size of the measurements for different masks are the same hence the same proportions of the $0$th order are acquired. Poisson and Gaussian noises are added to the Fourier intensity images in order to simulate the experimental environment. The Fourier intensity images are capped at the value $ 4095 $ (corresponding to $12$ bits) to simulate the dynamic range of the camera. We then use the TV-MAP algorithm in Eq. \eqref{TV-MAP} to retrieve the magnitude and phase of the original image.  To optimize the performance, we fine-tune the value of the regularization parameter $ \alpha $ in Eq.\eqref{TV-MAP} to ensure the best $ MSE_{phase} $ is obtained in each case. 
			
			The simulation results are shown in Table \ref{tablediffratio}. It can be seen that the $ SSE $ and $ MSE $ decrease as $  R_1 $ and $ \sigma $ increase up to a certain limit and then the $ SSE $ and $ MSE $ increase thereafter. We have explained in Section \ref{sec:analysis} why the high-frequency information of the mask will increase when $ R_1 $ increases beyond a certain limit and degrade the performance. Similar behavior is noted when $ \sigma $ increases. For binary amplitude masks,  increasing  $ \sigma $ will enlarge the dynamic range of the Fourier intensity image, which will introduce the saturation problem (also exists in the experiment) when $ \sigma $ increases up to a certain value and then affects the performance. In case that the binary phase masks are used, we have also noted in the experiments that the saturation problem occurs if $ \sigma  $ is set to be bigger than $ 0.6 $ or smaller than $ 0.4 $.  Consequently, the parameters $ R_1 $ and $ \sigma $ should be carefully chosen  in the experiments. It is indeed an advantage of using the FMEDg method for generating the green noise masks since it allows us to flexibly adjust the parameters $ R_1 $ and $ \sigma $ to achieve the required frequency characteristics and density to maximize the performance. 
			
			\begin{table*}[t]
				\centering
				\caption{Reconstruction performance with respect to different $\sigma$ and $ R_1 $ with green noise masks}
				\begin{tabular}{|c|c|c|c|c|}
					\hline
					\multirow{2}{*}{ \diagbox[width=1.8cm, height=1.25cm]{$\sigma$}{ $R_1$}} &  { $ 1.0 $ } &  { $ 1.5 $ } &  { $ 2.0 $ } &
					 { $ 2.5 $ } \\
					\cline { 2 - 5 }  & \makecell[c]{ $SSE_{amplitude}$ $(\mathrm{dB})$,\\ $MSE_{phase}$  } & \makecell[c]{$SSE_{amplitude}$ $(\mathrm{dB})$,\\ $MSE_{phase}$}  & \makecell[c]{$SSE_{amplitude}$ $(\mathrm{dB})$,\\ $MSE_{phase}$} & \makecell[c]{$SSE_{amplitude}$ $(\mathrm{dB})$,\\ $MSE_{phase}$}   \\
					\hline 
					$0.3$ & $ -11.943, 0.121 $ & $ -14.636, 0.089 $ & $ -13.467, 0.094 $ & $ -13.124, 0.094  $ \\ \hline
					$0.4$ & $ -13.116, 0.122 $ & $ -13.971, 0.079 $ & $ -14.527, 0.083  $ & $ -14.119, 0.090 $  \\ \hline
					$0.5$ & $ -13.880, 0.084  $ & $ \mathbf{-16.451, 0.054}  $ & $  -16.043, 0.058 $  & $ -15.795, 0.059 $ \\ \hline
					$0.6$ & $ -12.694, 0.116  $ & $  -14.527, 0.081 $  & $  -13.294, 0.082  $ & $ -14.757, 0.084 $ \\ \hline
					$0.7$ & $ -11.826, 0.241 $ & $ -13.745, 0.170 $ & $ -12.760, 0.175 $ &$  -12.373, 0.187 $  \\ \hline
				\end{tabular}
				\label{tablediffratio}
			\end{table*}

	\section{Experimental Results} \label{sec:4}

\begin{figure}[H]		
			\centering\includegraphics[width=0.5\linewidth]{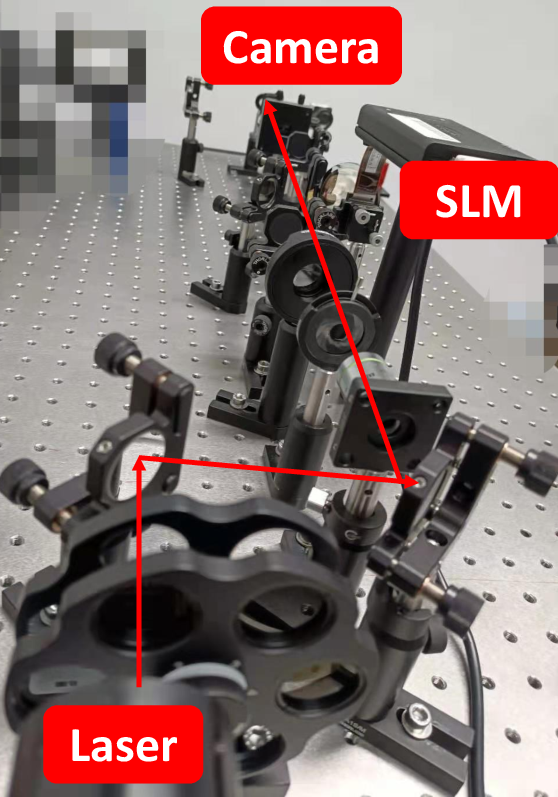}
			\caption{The hardware setup of the SLM-based CDP phase retrieval system.}
			\label{fig:SLMsetup}
		\end{figure}

		The aim of the experiments is to demonstrate the performance achieved by using the proposed green noise-masking scheme in an actual  CDP phase retrieval system. It is compared with  the traditional white noise masks \cite{candes2015code}. The experimental setup comprises a Thorlabs 10mW HeNe laser with wavelength $ \lambda = 632.8 $ nm; and a $ 12 $-bit $ 1920\times1200 $ Kiralux CMOS camera with pixel pitch $ 5.86 \mu $ m. For the implementation of the noise masks, a $ 1920\times1080 $ Holoeye Pluto phase-only SLM with pixel pitch $ \delta_{SLM}=8\mu m $ is used. Only the central $ 256\times256$ SLM pixels are utilized in all experiments.  The focal length of the lens behind the SLM is $ 75  mm$. As explained in Section {\ref{sec:2}}, the Nyquist frequency of the system is $ \frac{\lambda f}{2d} $ corresponding to the red square in Figure {\ref{fig:Fourierintensity}. For all experiments, we use the same set of optical devices hence the Nyquist frequency is the same as $ \frac{\lambda f}{2d} $. The data of the same rectangular areas are captured  for the phase retrieval algorithm using different kinds of random masks.} The whole experimental setup has already been shown in Section \ref{sec:2} (Figure \ref{fig:setup}).  Figure \ref{fig:SLMsetup} shows a photograph of the actual hardware setup.  
		
\begin{figure*}[t]		
			\centering\includegraphics[width=0.8\linewidth]{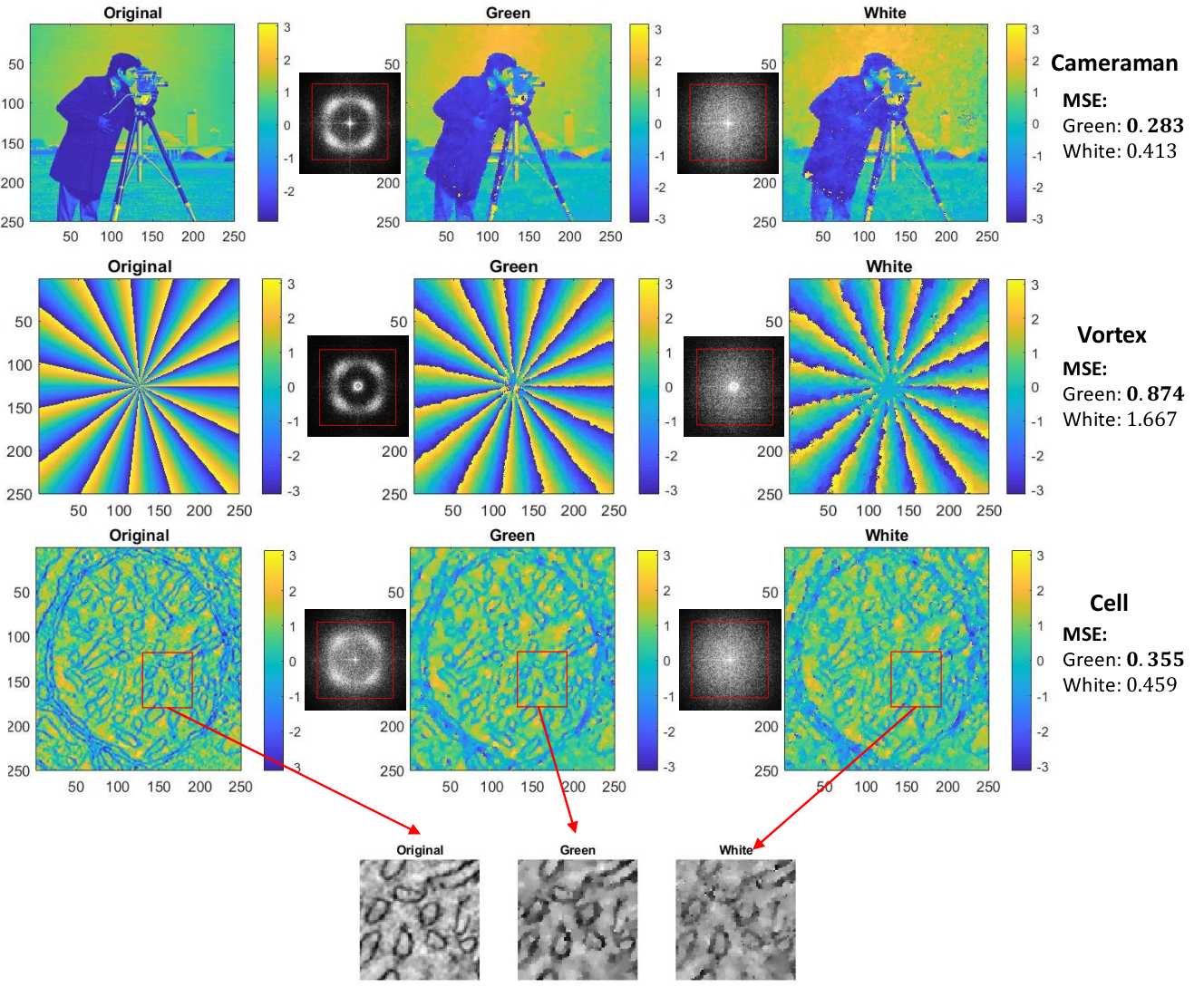}
			\caption{Experimental results  via the  green and white noise masks. Only the data inside red-square region will be considered in the phase retrieval algorithm.}
			\label{fig:SLMexp}
		\end{figure*}

		In the experiments, we also use the SLM to generate the testing objects. Specifically, three images namely, Cameraman, Vortex and Cell (see Figure \ref{fig:SLMexp})  are loaded to the SLM to generate the phase testing objects that can impose multiple-level phase changes. In this experiment, the binary phase masks $ \mathbf{I} \in\{1, -1\} $ are used. The images are pre-multiplied with the random masks before sending to the SLM. Therefore, the output of the SLM has already implemented the function $\mathbf{I}_i\circ\mathbf{U}$ in Eq. \eqref{CDP}. The green noise masks are generated using the FMEDg method described above. Based on the simulation result described in Section \ref{sec:choice} and further experimental work, we choose $R_1=1.5$ and $ \sigma=0.5 $ when generating the green noise masks. Only $ 3 $ measurements are taken for each kind of binary phase masks. The number of measurements $ M $ does not affect the conclusions of the experiment as long as it is not too small. If it is very small, the phase retrieval algorithm sometimes cannot converge. It is worth noting that the fixed constant $ \alpha $ in the optimization algorithm controls the strength of regularization. For different experimental objects, the optimal $ \alpha $ are  different because of their different complexities (curves, sizes, edges, etc.). In our experiments, we fine-tune the values of $ \alpha $ to obtain the minimum error in each case. The experimental results and the corresponding Fourier intensity measurements are presented in Figure \ref{fig:SLMexp}. It can be clearly seen that the performance of the reconstructed images via the green noise masks are much better than those reconstructed via the white noise masks, both quantitativel	y and qualitatively. The improvement in MSE can be up to $ 47.57\%$ . The improvement on the vortex images is particularly significant. A zoom-in region of the reconstructed cell image is presented to better visualize the performance of different masking schemes. It can be seen that textures and edges of the reconstructed cell image via the green noise masks are much clearer and sharper than those via the white noise masks.

\begin{figure*}[htbp]		
			\centering\includegraphics[width=0.7\linewidth]{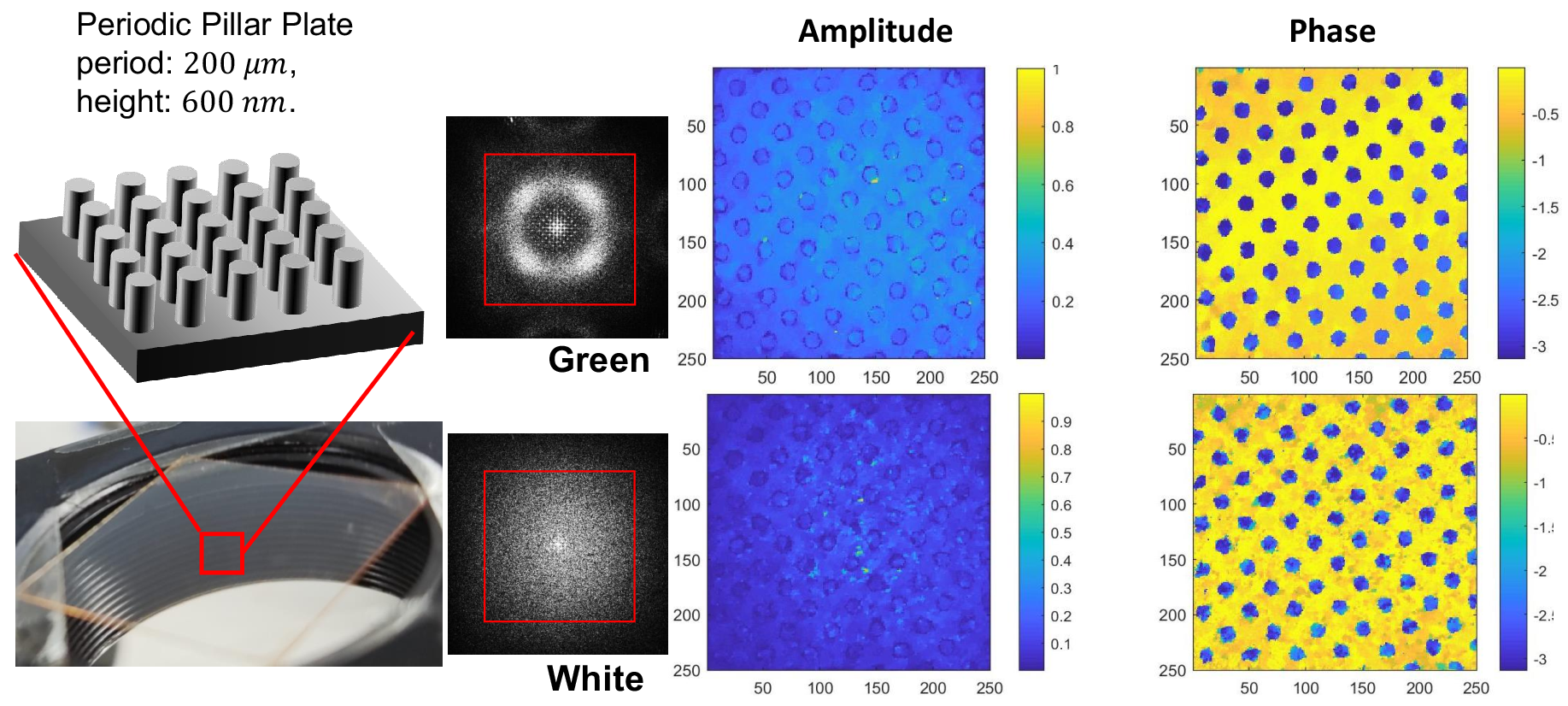}
			\caption{Experimental results of the periodic pillar grating with green and white random masks. $1$ measurement is used for each kind of masks. The image size is $250 \times 250$. Only the data inside red-square region will be considered by the phase retrieval algorithm.}
			\label{fig:plate}
		\end{figure*}

		We also use a real object which is a periodic pillar optical grating (shown in Figure {\ref{fig:plate}}) fabricated with SU-8 to demonstrate the effectiveness of the proposed green noise masks. The SU-8 is transparent under $632.8$ nm. The period of the grating is $200\ \mu m$ and the height is $d = 600$ nm. Hence the phase difference between air and SU-8 area is $\Delta \phi = \frac{2\pi d (n_{1} -n_{2})}{\lambda } \approx 0.98 \pi$, where $n_1 \approx 1.52$ and $n_2 = 1$ represent the refractive index of the SU-8 and air, respectively. We use the SLM in the last experiment to generate the same set of the phase-only random masks. The central $250 \times 250$  pixels of the SLM are used in the experiment. Only $1$ measurement is used for each kind of binary masks. Using more masks would not change the conclusion of the experiment.
	
		The experimental results are shown in Figure {\ref{fig:plate}}. For the amplitude part, periodic rings appear in the reconstructed images because of the sharp change along the edges of the pillars. As can be seen, the amplitude image via the green noise mask is smoother and more uniform than that of the white noise mask. For the phase part, the phase difference between the pillar areas and the air areas is approximately $0.98 \pi$ as stated in the above paragraph. It is clearly shown that the reconstructed image with the green noise mask can better reconstruct the phase part while there exists phase errors via the white noise mask. The phase errors mainly locate close to the edges of the pillar, which means that the lost high-frequency information degrades the performance. From the above experimental results, we have shown that the proposed green noise mask outperforms the traditional white noise mask both quantitatively and qualitatively when using in SLM-based phase retrieval systems. 
		
\begin{figure}[H]		
			\centering\includegraphics[width=\linewidth]{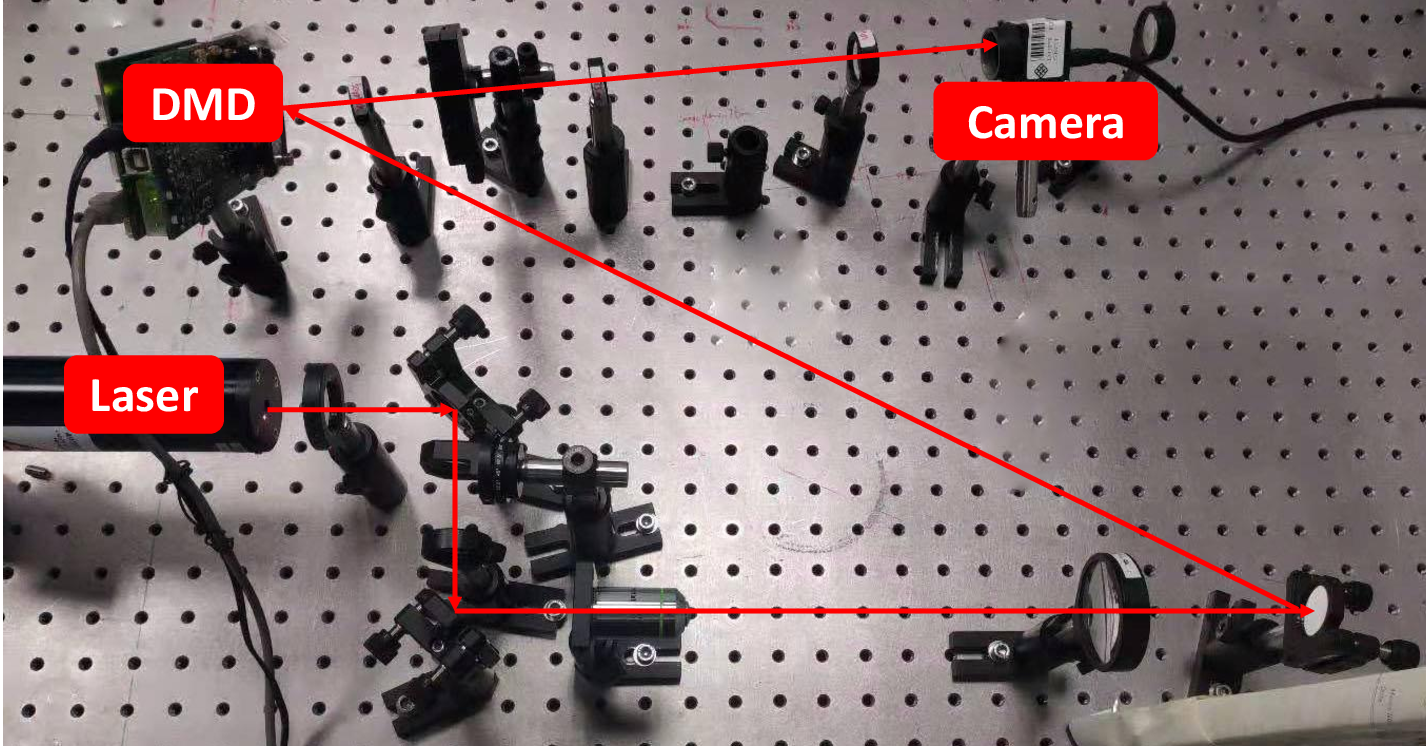}
			\caption{The actual DMD-based CDP phase retrieval system.}
			\label{fig:DMDsetup}
		\end{figure}

\begin{figure*}[t]		
			\centering\includegraphics[width=0.8\linewidth]{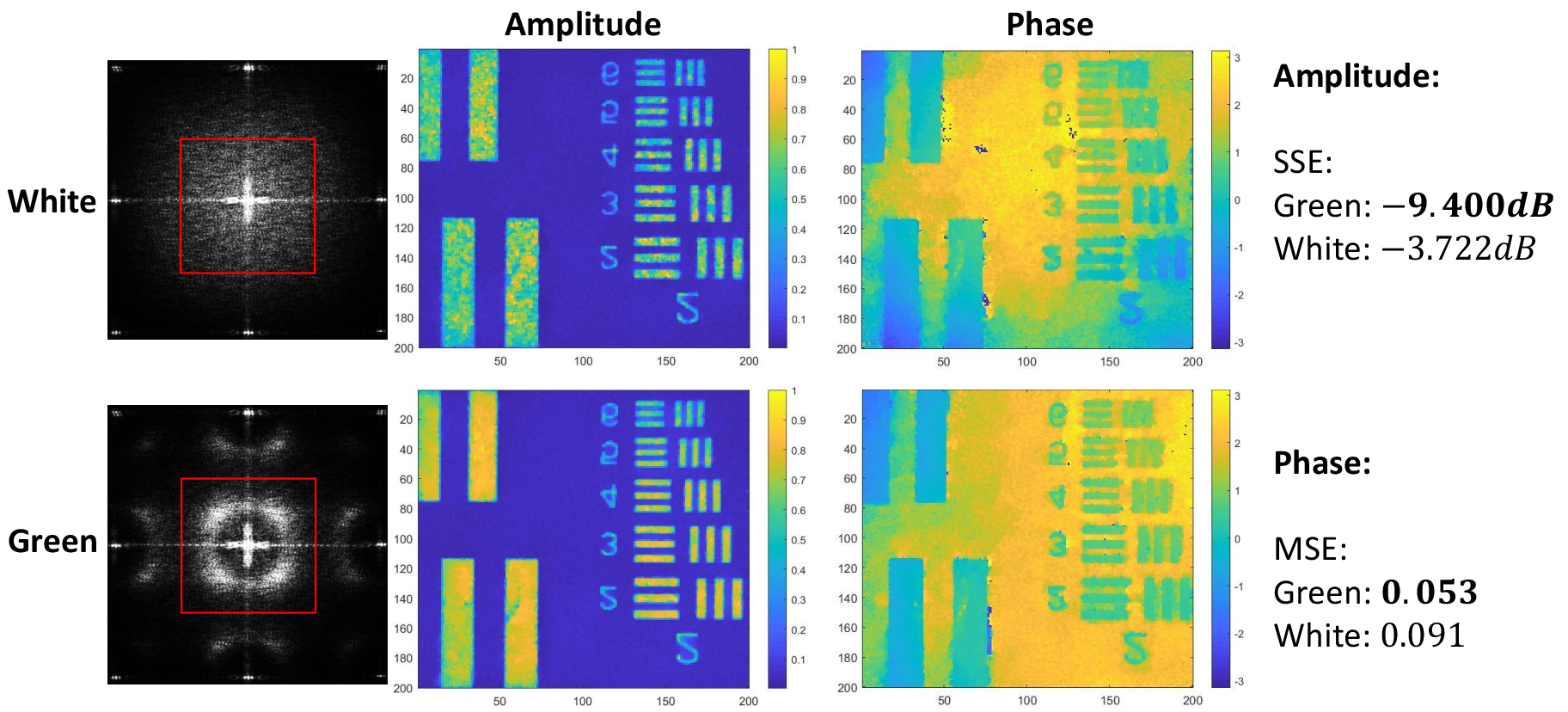}
			\caption{Experimental results of the USAF chart reconstructed via the green and white noise masks. Only the data inside red-square region will be considered by the phase retrieval algorithm.}
			\label{fig:USAF}
		\end{figure*}

		To demonstrate the generality of the proposed green noise masking scheme, we also apply the scheme to a DMD-based CDP phase retrieval system. Different from the experimental setup in Figure {\ref{fig:SLMsetup}}, a $ 1024\times768 $ pixels DMD with pixel pitch $ 13.68\mu m $ installed on a Discover $ 1100 $ DLP board is used to generate the random masks. The other parts of the setup can be found in Figure {\ref{fig:setup}}. Figure {\ref{fig:DMDsetup}} shows a photograph of the actual hardware setup. The optical path includes a reflective mirror (placed in the bottom-right corner) is used to match the blazed grating angle of the DMD board. Several reflective mirrors are also used in the laser beam magnification path because of the limited size of the optical table. In this experiment, the binary amplitude masks $ \mathbf{I} \in\{0, 1\} $ are loaded to the DMD board. A Newport USAF chart is placed on the object plane as the testing object. The projected light can only pass through the chart through its holes thus introduces sharp amplitude changes along the boundaries of the holes. The light that can pass through the chart should have the same phase, and we do not expect there is any variation in phase for the regions with no light. For the experiment, $ 3 $ measurements are captured by the camera for each kind of binary masks. The green noise masks are generated using the same $ R_1 $ and $ \sigma $ as in the SLM experiments. Similar to the SLM experiment, the number of measurements chosen does not affect the conclusions of the experiment as long as it is not too small. They are fed to the same TV-MAP algorithm to retrieve the amplitude and phase of the USAF chart. And we fine-tune the values of $\alpha $ to achieve the minimum error  in each case.  Figure \ref{fig:USAF} shows the experimental results and the corresponding Fourier intensity measurements with different binary masks. It can be seen that the amplitude and phase images retrieved through the use of different kinds of binary masks have clear differences. Even though the background areas of the amplitude images are smooth for all kinds of binary masks, the quality of the images in the regions where light can pass through has a considerable difference.  The amplitude image resulting from using the green noise masks remains sharp and relatively uniform. Comparing with the ground truth, the $ SSE (dB) $ of the amplitude images with the green and white noise masks are $ -9.400 $ dB and $ -3.722 $ dB, respectively. As to the phase images, the one reconstructed with the green noise masks has a much smoother background and fewer errors compared with the white noise masks.
		
	\section{Conclusion}
	
		This paper proposes a green noise binary masking scheme for the  coded diffraction pattern (CDP) phase retrieval systems. We have explained why the high-frequency data in the Fourier intensity measurements will be ignored in the CDP phase retrieval process. To reduce the information loss, we suggested using  green noise masks that have very low high-frequency concentrations. The green noise masks are designed using a digital halftoning method we developed earlier. It has the advantage that the principal frequency and the density of the noise mask can be controlled independently. Also, we have demonstrated how the high-frequency data in the Fourier intensity measurements can affect the quality of the reconstruction images. It is verified by the experimental results, which show that the quality of the amplitude and phase images retrieved using the proposed green noise binary masking scheme outperforms the commonly used white noise binary masks when applying to two CDP phase retrieval systems.   
		
	\section*{Funding} This work was supported by the Hong Kong Research Grant Council under General Research Fund no. PolyU 152478/16E.

	\section*{Declaration of Competing Interest} The authors declare no conflicts of interest.
	
	\section*{CRediT authorship contribution statement} \textbf{Qiuliang Ye}: Conceptualization, Methodology, Simulation, Experiment, Writing - original draft, Writing - review \& editing.  \textbf{Yuk-Hee Chan}: Conceptualization, Methodology.   \textbf{Michael G. Somekh}: Supervision, Writing - review \& editing. \textbf{Daniel P.K. Lun}: Supervision, Project administration, Funding acquisition, Conceptualization, Methodology, Writing - review \& editing.

	\bibliography{Reference_All}

\end{document}